\documentclass[a4paper,10pt]{article}

\usepackage{subfig}
\usepackage{graphicx}
\usepackage{cite}
\usepackage{epsfig,amsmath,amssymb,verbatim,mathrsfs,array,layout,textcomp,amssymb,latexsym}

\newcommand{\beq}{\begin{eqnarray}}
\newcommand{\eeq}{\end{eqnarray}}
\newcommand{\be}{\begin{eqnarray}}
\newcommand{\ee}{\end{eqnarray}}

\def\beqa{\begin{eqnarray}}
\def\eeqa{\end{eqnarray}}
\def\bea{\begin{eqnarray}}
\def\eea{\end{eqnarray}}
\newcommand{\no}{\nonumber}
\newcommand{\bv}{\left(\begin{array}{c}}
\newcommand{\ev}{\end{array}\right)}
\newcommand{\bmtwo}{\left(\begin{array}{cc}}
\newcommand{\bmthree}{\left(\begin{array}{ccc}}
\newcommand{\emn}{\end{array}\right)}
\newcommand{\bmtwoc}{\left\{\begin{array}{cc}}
\newcommand{\bmthreec}{\left\{\begin{array}{ccc}}
\newcommand{\emnc}{\end{array}\right\}}
\newcommand{\ba}{\begin{array}}
\newcommand{\ea}{\end{array}}

\def\lsim{\mathrel{\rlap{\lower4pt\hbox{\hskip1pt$\sim$}}
     \raise1pt\hbox{$<$}}}         
\def\gsim{\mathrel{\rlap{\lower4pt\hbox{\hskip1pt$\sim$}}
     \raise1pt\hbox{$>$}}}         

\usepackage{xcolor}

\usepackage[nolist]{acronym}
\begin{acronym}
	\acro{SM}{Standard Model}
	\acro{CPV}{CP violating}
	\acro{CPC}{CP conserving}
	\acro{EH}{Euler-Heisenberg}
	\acro{BSM}{beyond the Standard Model}
	\acro{COM}{center of mass}
	\acro{QED}{Quantum Electrodynamics}
	\acro{EFT}{Effective Field Theory}
	\acro{ALP}{axion-like particle}
	\acro{EOM}{Equation of Motion}
	\acro{RHS}{right hand side}
	\acro{LHS}{left hand side}
	\acro{FP}{Fabry-Perot}
	\acro{EM}{Electromagnetism}
	\acro{SRF}{superconducting radio frequency}
	\acro{SNR}{signal to noise ratio}
	\acro{EP}{Equivalence Principle}
	\acro{CL}{confidence level}
	\acro{DM}{dark matter}
	\acro{ULDM}{ultra-light dark matter}
	\acro{VEV}{vacuum expectation value}
	\acro{EWSB}{Electroweak Symmetry Breaking}
\end{acronym}

\begin{document}

\begin{titlepage}

\vskip1.5cm
\begin{center}
{\Large \bf Parametric resonance in neutrino oscillations induced by ultra-light dark matter and implications for KamLAND and JUNO} \\
\end{center}
\vskip0.2cm

\begin{center}
Marta Losada$^1$, Yosef Nir$^2$,  Gilad Perez$^2$, Inbar Savoray$^2$ and Yogev Shpilman$^2$\\
\end{center}
\vskip 8pt

\begin{center}
{ \it $^1$New York University Abu Dhabi, PO Box 129188, Saadiyat Island, Abu Dhabi, United Arab Emirates\\
$^2$Department of Particle Physics and Astrophysics,\\
Weizmann Institute of Science, Rehovot 7610001, Israel} \vspace*{0.3cm}

{\tt   marta.losada@nyu.edu, yosef.nir,gilad.perez,inbar.savoray, yogev.shpilman@weizmann.ac.il}
\end{center}

\vglue 0.3truecm

\begin{abstract}
  \noindent
  If \ac{ULDM} exists and couples to neutrinos, the neutrino oscillation probability might be significantly altered by a parametric resonance. This resonance can occur if the typical frequency of neutrino flavor-oscillations $\Delta m^2/(2E)$, where $\Delta m^2$ is the mass-squared difference of the neutrinos and $E$ is the neutrino energy, matches the oscillation frequency of the \ac{ULDM} field, determined by its mass, $m_\phi$.
The resonance could lead to observable effects even if the \ac{ULDM} coupling is very small, and even if its typical oscillation period, given by $\tau_\phi=2\pi/m_\phi$, is much shorter than the experimental temporal resolution.
Defining a small parameter $\epsilon_\phi$ to be the ratio between the contribution of the \ac{ULDM} field to the neutrino mass and the vacuum value of the neutrino mass, the impact of the resonance is particularly significant if $\epsilon_\phi m_\phi L\gsim 4$, where $L$ is the distance between the neutrino source and the detector. Such parametric resonance can improve the fit to the KamLAND experiment measurements by about $3.5\sigma$ compared to standard oscillations. This scenario will be tested by the JUNO experiment.
\end{abstract}

\end{titlepage}

\acresetall
\section{Introduction}
Light scalar fields may constitute \ac{ULDM} candidates \cite{Arvanitaki:2014faa, Graham:2015ifn,Stadnik:2014tta,Banerjee:2018xmn}. If these fields are sufficiently light, they oscillate in time, and can lead to temporal variations in various constants of Nature. We are interested in the possibility that the \ac{ULDM} field couples to neutrinos, thus affecting their masses and mixing \cite{Krnjaic:2017zlz,Brdar:2017kbt,Capozzi:2018bps,Berlin:2016woy,Dev:2020kgz,Losada:2021bxx,Chun:2021ief,Huang:2021kam}.

A scalar with mass $m_\phi\lsim{\rm eV}$ can be treated as a classical bosonic field that oscillates with time,
\beq
\label{eq:scalarOscillations}\phi=\phi_0\sin(m_\phi t+\varphi)\,,
\eeq
with initial phase $\varphi$. If it is to account for the \ac{DM}, then
\beq
\phi_0=\frac{\sqrt{2\rho_\phi^\oplus}}{m_\phi}\sim2\ {\rm GeV}\ \left(\frac{10^{-12}\ {\rm eV}}{m_\phi}\right),
\eeq
where $\rho_\phi^\oplus$ corresponds to the \ac{ULDM} density at the surface of Earth, which we assume to match the \ac{DM} density at the solar position given in~\cite{Salucci:2010qr}. The oscillation period is given by
\beq
\tau_\phi=\frac{2\pi}{m_\phi}\approx4\times10^{-3}\ {\rm sec}\ \left(\frac{10^{-12}\ {\rm eV}}{m_\phi}\right)\,.
\eeq
In a previous work \cite{Losada:2021bxx}, we considered the range of $m_\phi\lsim10^{-12}\ {\rm eV}$. For this mass range, $\tau_\phi$ is larger than $\tau_d(=L)$, the source-to-detector distance, in which case the effective neutrino masses and mixing do not change along the propagation from the source to the detector. In this work, we consider the range 
\beq
m_\phi\gsim10^{-12}\ {\rm eV},
\eeq
leading to $\tau_\phi<\tau_d$. This case opens the door to the possibility of parametric resonance, which is the main focus of this paper.

Consider the effective mass and $\phi$-Yukawa terms for the neutrinos, in the neutrino mass basis, arising from dimension-five and dimension-six terms in the Lagrangian, respectively:
\beq
{\cal L}_{m_\nu}=m_i \nu_i^T\nu_i + \hat y_{ij}\phi\nu_i^T\nu_j\,.
\eeq
Treating $\phi$ as a classical field, it modifies the neutrino mass matrix as
\beq
(\hat m_\nu)_{ij}=m_i\delta_{ij}+\hat y_{ij}\phi\,.
\eeq
We are mainly interested in the off-diagonal $\hat y$ entries. We assume that the $\phi$-dependent modifications are small, and define small parameters $\epsilon_{ij}$ given by
\beq\label{eq:epsilonij}
\epsilon_{ij}\equiv\frac{2\hat y_{ij}\phi_0}{m_i-m_j}.
\eeq
For the two generation case discussed below, we denote the small parameter by $\epsilon_\phi$.
In what follows, we show that a parametric resonance, that is an ${\cal O}(1)$ modification of the neutrino oscillation probability, can occur even for very small $\epsilon_{ij}$, as their effects are enhanced in the $m_\phi L\gg1$ limit. Such large contributions are generated when the oscillation frequency of the neutrinos,
\beq
\omega_E\equiv\Delta m^2_{ij}/(4E),
\eeq
(where $\Delta m^2_{ij}\equiv m_i^2-m_j^2$), matches half the oscillation frequency of the scalar field,
\beq
\omega_R\equiv m_\phi/2.
\eeq

The plan of this paper is as follows. In Section \ref{sec:theory}, we present our theoretical framework, develop our formalism, and explain the conditions for and features of the parametric resonance. In Section \ref{sec:exp}, we study the implications for neutrino experiments. We argue that a feature in the spectrum measured by KamLAND, if not a statistical fluctuation, can be accounted for by a parametric resonance due to a \ac{ULDM} field, and moreover that the JUNO experiment will test this scenario. We summarize our conclusions in Section \ref{sec:con}.
 
\section{The theoretical framework}
\label{sec:theory}
\subsection{Two neutrino model}
To understand the basic features of the \ac{ULDM}-induced parametric resonance in neutrino oscillations, we first study a two neutrino generations model. Since we are interested in a resonance effect, the periodic perturbation induced by the \ac{ULDM} field should couple the two states of the system. We thus consider the following Yukawa matrix in the neutrino mass basis:
\beq
\label{eq:2nuYMatrix}\hat y=y\begin{pmatrix} 0 & 1\cr 1 & 0 \end{pmatrix},
\eeq
with $y$ real. Two comments are in order:
\begin{itemize}
\item Diagonal entries in $\hat y$ cause an effect of ${\cal O}(\epsilon_\phi^2)$ in the time-averaged neutrino oscillation probability. (By ``time-averaged" we mean an average of the oscillation probability over the initial phase $\varphi$ of the scalar field in Eq.~(\ref{eq:scalarOscillations}).) Since this is a minuscule effect compared to the potential ${\cal O}(1)$ effect of the resonance, we set the diagonal $\hat y$ terms to zero.
\item If $y$ were complex, we could absorb its phase in the phase $\varphi$ (see Appendix~\ref{appendix:complex}). It will then have no effect on the time averaged transition probability. 
\end{itemize}
Considering the above, and the fact that $\hat{y}$ is symmetric, the matrix in Eq.~(\ref{eq:2nuYMatrix}) is the most general matrix needed for this study.

The equations of motion for the neutrinos are given by
\beq
i\partial_t\begin{pmatrix}\nu_\alpha\cr \nu_\beta\end{pmatrix}={\cal H}\begin{pmatrix}\nu_\alpha\cr \nu_\beta\end{pmatrix},
\eeq
where the effective Hamiltonian in the unperturbed mass basis is given by
\beq
{\cal H}=\frac{1}{2E}(m_\nu+\hat y\phi)^\dagger(m_\nu+\hat y\phi).
\eeq
For our choice of $\hat y$, and omitting terms proportional to the unit matrix, which do not affect neutrino oscillations, we obtain
\beq
{\cal H}=\omega_E[\sigma_z+\epsilon_\phi\sin(m_\phi t+\varphi)\sigma_x],
\eeq
where $\sigma$ are the Pauli matrices. 

We label the unperturbed neutrino mass eigenstates as $\nu^0_{1,2}$. Their time evolution is given by
\beq \label{eq:nu0sol}
\nu^0_{1,2}(t)=e^{\mp i\omega_E t}\nu^0_{1,2}(t=0).
\eeq
In the presence of the perturbation, the time evolution is given by
\beq\label{eq:defc}
\nu_{1,2}(t)= c_{1,2}(t)\nu^0_{1,2}(t),
\eeq
where $c_{1,2}$ obey the following equations:
\beq\label{eq:coom}
i\partial_t c_{1,2}=\frac{\omega_E}{2i}\epsilon_\phi\left[e^{i(m_\phi\pm2\omega_E)t+i\varphi}-e^{-i(m_\phi\mp2\omega_E)t-i\varphi}\right]c_{2,1}.
\eeq
We further define a detuning parameter,
\beq \label{eq:defdeltaE}
\delta_E\equiv\frac{m_\phi-2\omega_E}{m_\phi}=\frac{\omega_R-\omega_E}{\omega_R}\,,
\eeq
and rewrite Eq.~(\ref{eq:coom}) in terms of $\delta_E$:
\beq
i\partial_t c_{1,2}=\frac{\omega_E}{2i}\epsilon_\phi\left\{e^{im_\phi[1\pm(1-\delta_E)]t+i\varphi}-e^{-im_\phi[1\mp(1-\delta_E)]t-i\varphi}\right\}c_{2,1}.
\eeq
For $\delta_E\ll1$, we find a term that oscillates with frequency $\delta_E m_\phi$ and a term that oscillates with frequency $2m_\phi$. If the typical time scale for neutrino propagation satisfies
\beq\label{eq:taudtauphi}
\tau_d\gg \tau_\phi,
\eeq
then the rapidly oscillating term can be neglected. This is known as the ``rotating wave approximation", and it leads to a simplified equation of motion 
\beq
i\partial_t c_{1,2}=\mp \frac{\omega_E}{2i}\epsilon_\phi  e^{\mp i\left(m_\phi\delta_E t +\varphi\right)} c_{2,1}\,,
\eeq
that can be solved analytically. Then, in the vicinity of the resonance, we obtain the probability for an interaction eigenstate $\nu_\beta$ produced at time $t$ to be measured as the interaction eigenstate $\nu_{\alpha}$ after propagating a distance $L$,
\beqa
P_{\alpha\beta}(t)&=&\frac{1}{\epsilon_\phi^2+4\delta_E^2}\left\{\sin\left(\frac{\omega_E L}{2}\sqrt{\epsilon_\phi^2+4\delta_E^2}\right)
\left[\epsilon_\phi c_{2\theta}\sin\left(\varphi+2\omega_Rt+\omega_R L\right)-2\delta_E s_{2\theta}\cos(\omega_R L)\right]\right.\no\\
&+&\left.\sqrt{\epsilon_\phi^2+4\delta_E^2} s_{2\theta}\sin(\omega_R L)\cos\left(\frac{\omega_E L}{2}\sqrt{\epsilon_\phi^2+4\delta_E^2}\right)\right\}^2\no\\
&+&\frac{\epsilon_\phi^2}{\epsilon_\phi^2+4\delta_E^2}\cos^2\left(\varphi+2\omega_Rt+\omega_R L\right)\sin^2\left(\frac{\omega_E L}{2}\sqrt{\epsilon_\phi^2+4\delta_E^2}\right),
\eeqa
where $s_{2\theta}\equiv\sin2\theta$, $c_{2\theta}\equiv\cos2\theta$, and $\theta$ is the rotation angle from the interaction basis (that is also the charged lepton mass basis) to the unperturbed neutrino mass basis. When deriving the transition probability measured by experiments, where temporal dependence is not considered, we have to average the above expression over the initial \ac{ULDM} phase $\varphi$. This would also be the case if the events are time-stamped, but the \ac{ULDM} oscillation time or coherence time are shorter than the experimental resolution.

The transition probability exhibits a resonance at $\omega_E=\omega_R$, which could lead to $\mathcal{O}\left(1\right)$ modifications to the standard transition probability when $m_\phi L\epsilon_\phi\gsim4$. This resonance is analogous to the parametric resonance in classical mechanics, which is obtained when the frequency of the perturbation is twice the oscillator frequency. Note that our system has two small parameters: $\epsilon_\phi$ and $(m_\phi L)^{-1}$, and the transition probability depends on the ratio $\epsilon_\phi/(m_\phi L)^{-1}$, which could be much larger than $\epsilon_\phi$. 

The resonance corresponds to a situation where each interaction of the neutrinos with the scalar results in an exchange of the neutrino mass eigenstate, with corresponding absorption or emission of energy quanta $m_\phi$. The $\nu_1\leftrightarrow\nu_2$ transition would be most efficient when energy conservation is satisfied. This is possible for any $N=2n+1$ interactions with the scalar, leading to a resonance at $\omega_E=(N/2)m_\phi$. However, a large $N$ resonance comes at the cost of suppressed effective coupling $\epsilon_N$ \cite{Shirley:1965}:
\beq\label{eq:epsilonN}
\frac{\epsilon_N}{\epsilon_\phi}\approx\frac{(\epsilon_\phi N)^{N-1}}{2^{2N-2}(\frac{N-1}{2}!)^2}.
\eeq
%

\subsubsection{$\theta=0$}
Various features of the resonance are simplest to obtain in a two generation model with zero mixing, $\theta=0$. With the initial condition
\beq
\begin{pmatrix}\nu_\alpha \cr \nu_\beta\end{pmatrix}(t=0)=\begin{pmatrix} 0 \cr 1\end{pmatrix},
\eeq
the transition probability in vacuum is zero, as the neutrinos do not mix. The interaction with the \ac{ULDM} field, however, does allow transitions. 
For $\theta=0$, we obtain the well-known Rabi oscillations probability\footnote{Rabi oscillations would occur also for oscillatory matter profiles, see Ref. \cite{Ma:2018key} and references therein.}:
\beq
\langle P_{\alpha\beta}\rangle_\varphi=
\frac{\epsilon_\phi^2(1-\delta_E)^2}{\epsilon_\phi^2(1-\delta_E)^2+4\delta_E^2}\sin^2\left(\frac{m_\phi L}{4}\sqrt{\epsilon_\phi^2(1-\delta_E)^2+4\delta_E^2}\right).
\eeq

First, we examine the height of the resonance. At the resonance energy $E_R=\Delta m^2/(2m_\phi)$, corresponding to $\omega_E=m_\phi/2$ and $\delta_E=0$, we have for the phase-averaged transition probability,
\beq
P_{\alpha\beta}(E_R)=\sin^2\left(\frac{\epsilon_\phi m_\phi L}{4}\right).
\eeq
(Here and below, we often omit the $\langle\ldots\rangle_\phi$ notation for $P_{\alpha\beta}$.)
We can distinguish two possible regions:
\begin{itemize}
\item For $m_\phi L\epsilon_\phi/4\gsim1$, an ${\cal O}(1)$ transition probability might be induced, in contrast to the zero probability in vacuum.
\item For $m_\phi L\epsilon_\phi/4\lsim1$, we have $P_{\alpha\beta}(E_R)\approx(\epsilon_\phi m_\phi L/4)^2$, so that even for very small $\epsilon_\phi$, the effect can be enhanced by $m_\phi L\gg1$.
\end{itemize}

Next, we examine the shape of the probability around the resonance. Note that moving away from the resonance to higher energies, $E> E_R$, implies $0\leq \delta_E\leq1$, while moving away to lower energies, $E<E_R$, implies $-\infty\leq\delta_E<0$. We distinguish two limits:
\beq
P_{\alpha\beta}\approx\left\{\begin{matrix}
\frac{\epsilon_\phi^2(1-\delta_E)^2m_\phi^2L^2}{16} & \frac{m_\phi L}{4}\sqrt{4\delta_E^2+\epsilon_\phi^2(1-\delta_E)^2}\ll1 \cr
\frac{\epsilon_\phi^2}{\epsilon_\phi^2+(\frac{2\delta_E}{1-\delta_E})^2} & \frac{m_\phi L}{4}\sqrt{4\delta_E^2+\epsilon_\phi^2(1-\delta_E)^2}\gg1
\end{matrix}\right. .
\eeq
We obtain the width of the resonance:
\beq
\frac{|\delta E|}{E_R}=\frac{|\delta\omega_E|}{|\omega_R|}\sim
\left\{\begin{matrix}
\epsilon_\phi & \epsilon_\phi m_\phi L\gg1 \cr
1/(m_\phi L) & \epsilon_\phi m_\phi L\ll1
\end{matrix}\right. .
\eeq

\begin{figure}[ht]
 \begin{center}
\includegraphics[width=0.46\textwidth]{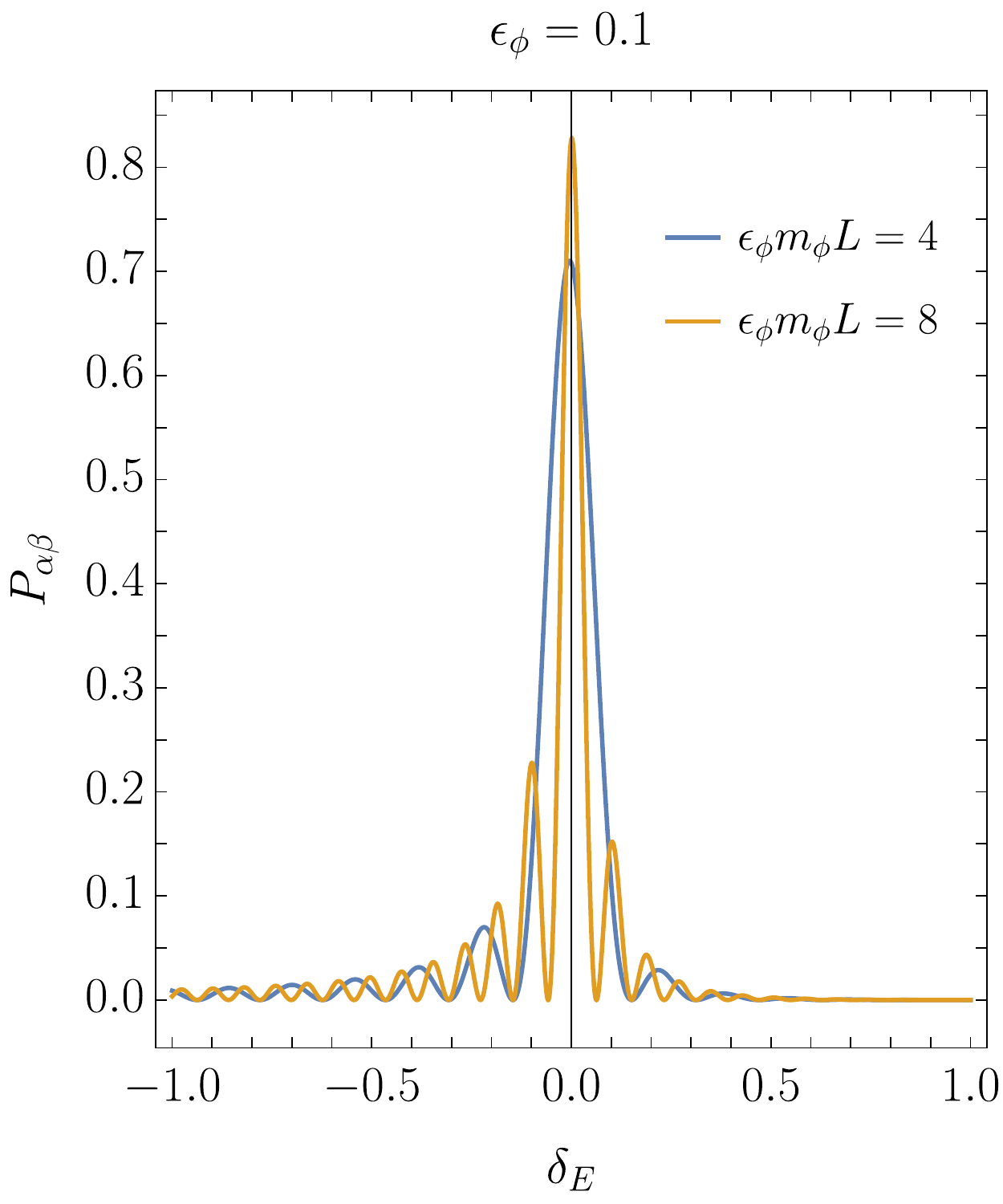}
\includegraphics[width=0.46\textwidth]{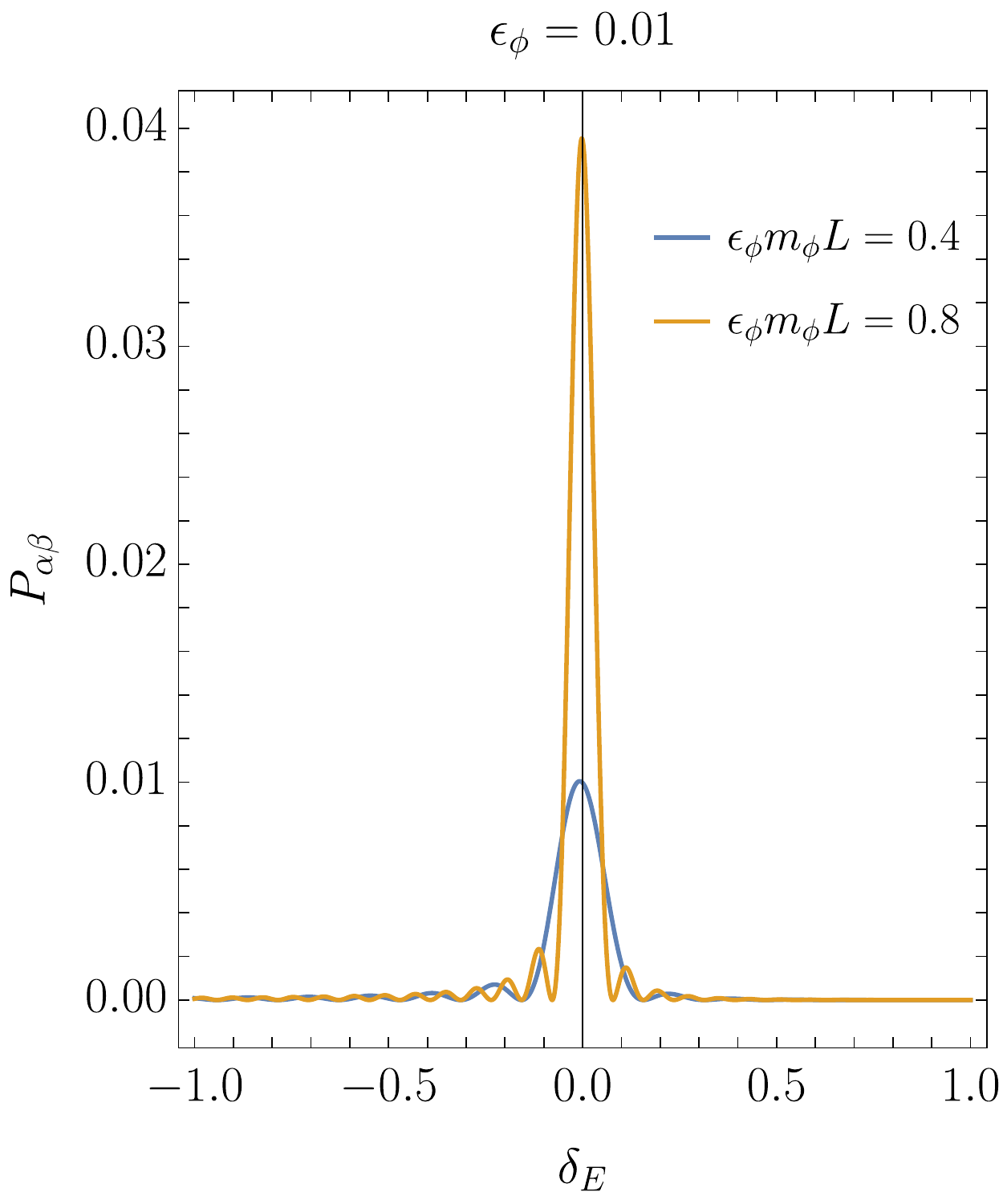}
  \caption{$P_{\alpha\beta}(\delta_E)$: \textit{(Left)}  $\epsilon_\phi m_\phi L\gg1$: The height of the resonance is ${\cal O}(1)$, and the width is $\epsilon_\phi$. \textit{(Right)} $\epsilon_\phi m_\phi L\ll1$: The height of the resonance is $(\epsilon_\phi m_\phi L)^2/16$, and the width is $1/(m_\phi L)$.
  }
  \label{fig:peml}
 \end{center}
\end{figure}

The features of the resonance are demonstrated in Fig.~\ref{fig:peml}.

\subsubsection{$\theta\neq0$}
Upon averaging over the initial phase $\varphi$, we obtain 
{\small
\beqa\label{eq:pabave}
\langle P_{\alpha\beta}\rangle_\varphi
&=&\frac{s_{2\theta}^2}{\epsilon_\phi^2+4\delta_E^2}\left[-2\delta_E\sin\left(\frac{\omega_E L}{2}\sqrt{\epsilon_\phi^2+4\delta_E^2}\right)\cos(\omega_R L)
+\sqrt{\epsilon_\phi^2+4\delta_E^2} \sin(\omega_R L)\cos\left(\frac{\omega_E L}{2}\sqrt{\epsilon_\phi^2+4\delta_E^2}\right)\right]^2\no\\
&+&\frac{\epsilon_\phi^2}{\epsilon_\phi^2+4\delta_E^2}\frac{1+c_{2\theta}^2}{2}\sin^2\left(\frac{\omega_E L}{2}\sqrt{\epsilon_\phi^2+4\delta_E^2}\right).
\eeqa
}
Given that the vacuum transition probability at $E=E_R$ is given by
\beq
P^0_{\alpha\beta}(E_R)=s_{2\theta}^2\sin^2(\omega_R L),
\eeq
we obtain:
\beq
P_{\alpha\beta}(E_R)=P^0_{\alpha\beta}(E_R)+\sin^2\left(\frac{\epsilon_\phi m_\phi L}{4}\right)\left[1-\frac12 s_{2\theta}^2-P^0_{\alpha\beta}(E_R)\right].
\eeq
Thus, the transition probability at the resonance is enhanced compared to the vacuum transition probability if $\sin^2(m_\phi L/2)<(1/s_{2\theta}^2)-(1/2)$ and reduced if $\sin^2(m_\phi L/2)>(1/s_{2\theta}^2)-(1/2)$.

\subsection{Three neutrino model}
\label{sec:3nu}
We consider the three neutrinos case, where the $\phi$-field couples off-diagonally to only two mass eigenstates. Explicitly, we take $\hat y$ in the neutrino vacuum mass basis of the following form: 
\beq\label{eq:3numodel}
\hat y=y\begin{pmatrix} 0 & 0 & 1 \cr 0 & 0 & 0 \cr 1 & 0 & 0 \end{pmatrix}\,.
\eeq
In this model, $\nu_2$ evolves independently of $\nu_1$ and $\nu_3$. Thus, the former approximately follows vacuum evolution, while for the latter two we can apply our two generations analysis. The final result for the $\varphi$-averaged survival probability is given by
{\small
\begin{align}
\langle P_{\alpha\alpha}\rangle_\varphi
&=\left\{2(|U_{\alpha1}|^2+|U_{\alpha3}|^2)\left[s^2_\epsilon\cos\left(\frac{m_\phi(1-\eta_E)}{2}L\right)+c^2_\epsilon\cos\left(\frac{m_\phi(1+\eta_E)}{2}L\right)\right]
+|U_{\alpha2}|^2\cos\left(\frac{(\Delta m^2_{21}+\Delta m^2_{23}) L}{4E}\right)\right\}^2\no\\
&+\left\{2(|U_{\alpha1}|^2-|U_{\alpha3}|^2)\left[c^2_\epsilon\cos\left(\frac{m_\phi(1-\eta_E)}{2}L\right)-s^2_\epsilon\cos\left(\frac{m_\phi(1+\eta_E)}{2}L\right)\right]
+|U_{\alpha2}|^2\sin\left(\frac{(\Delta m^2_{21}+\Delta m^2_{23}) L}{4E}\right)\right\}^2\no\\
&+8|U_{\alpha1}|^2|U_{\alpha3}|^2s^2_{\epsilon}c^2_\epsilon\sin^2\left(\frac{m_\phi\eta_E}{2}L\right),
\end{align}
}
where $s_\epsilon\equiv\sin\theta_\epsilon$,  $c_\epsilon\equiv\cos\theta_\epsilon$,
\beqa
\tan2\theta_\epsilon&=&-\frac{\epsilon_{31}(1-\delta_E)}{2\delta_E},\no\\
\eta_E&=&\frac12\sqrt{\epsilon_{31}^2(1-\delta_E)^2+4\delta_E^2}\,,
\eeqa
and $U_{\alpha i}$ are the entries of the PMNS matrix, which is the transition matrix from the unperturbed neutrino mass basis to the interaction basis.

At the resonance neutrino energy $E_R=\Delta m^2_{31}/(2m_\phi)$, implying $\sin2\theta_\epsilon=-1$ and $\eta_E=\epsilon_{31}/2$, the survival probability is given by
\beqa
P_{\alpha\alpha}(E_R)
&=&\left[2(|U_{\alpha1}|^2+|U_{\alpha3}|^2)\cos\frac{\epsilon_{31}m_\phi L}{4}\cos\frac{m_\phi L}{2}
+|U_{\alpha2}|^2\cos\frac{(\Delta m^2_{21}+\Delta m^2_{23}) L}{4E}\right]^2\no\\
&+&\left[2(|U_{\alpha1}|^2-|U_{\alpha3}|^2)\sin\frac{\epsilon_{31}m_\phi L}{4}\sin\frac{m_\phi L}{2}
+|U_{\alpha2}|^2\sin\frac{(\Delta m^2_{21}+\Delta m^2_{23}) L}{4E}\right]^2\no\\
&+&2|U_{\alpha1}|^2|U_{\alpha3}|^2\sin^2\frac{\epsilon_{31}m_\phi L}{4}.
\eeqa

Our main interest concerns $P_{ee}$. Making the approximation $|U_{e3}|=0$, we find, at the resonance,
\beqa
P_{ee}(E_R)&\approx&1-\sin^22\theta_{12}\sin^2\left(\frac{\Delta m^2_{21}t}{4E}\right)\no\\
&-&\left[\cos^4\theta_{12}\sin^2\left(\frac{\epsilon_{31}\Delta m^2_{31}t}{8E}\right)+\sin^22\theta_{12}\cos\left(\frac{\Delta m^2_{21}t}{2E}\right)\sin^2\left(\frac{\epsilon_{31}\Delta m^2_{31}t}{16E}\right)\right].
\eeqa

\section{Implications for experiments}
\label{sec:exp}
In order to obtain good sensitivity to the neutrino oscillation parameters, experiments where both the source and the detector are terrestrial are designed with energy spectrum and baseline length such that $\Delta m^2_{ij}L/(4E)={\cal O}(1)$ for some $\Delta m^2_{ij}$. A \ac{ULDM} field coupling these $\nu_i$ and $\nu_j$, and which is of mass $m_\phi=\Delta m^2_{ij}/(2E_R)$, where $E_R$ falls within the spectrum of the experiment, will not undergo many oscillations during the neutrino propagation. In this case, an $\mathcal{O}(1)$ modification of the oscillation probability would require $\epsilon_{ij}\sim 1/(m_\phi L)\geq\mathcal{O}(1)$. Furthermore, the width of the resonance, $\Delta E/E_R\gtrsim 1/(m_\phi L)$ will be at least of ${\cal O}(1)$, and thus spread over many energy bins (assuming energy resolution of a few percent). Given measurements of a neutrino oscillation experiment, one can derive bounds on the parameter space in such scenario, which we leave for future work. 

Nature has provided us, however, with a hierarchy of neutrino mass-squared differences, $\Delta m^2_{21}\ll\Delta m^2_{31}\sim\Delta m^2_{32}$. Consequently, a \ac{ULDM} field of mass $m_\phi=\Delta m^2_{31}/(2E_R)$ would oscillate many times during the neutrino propagation in an experiment where $\Delta m^2_{21}/(4E_R)={\cal O}(1)$, such that a significant resonance effect can be generated with $\epsilon_{13}\ll1$. Furthermore, the width of the resonance would be rather narrow,
\beq
\frac{\Delta E}{E_R}\sim\frac{2E_R}{\Delta m^2_{31}L}
\sim\frac{\Delta m^2_{21}}{\Delta m^2_{31}}\approx0.03,
\eeq
thus affecting a single, or very few adjacent energy bins.

We are thus led to consider, on the experimental side, the KamLAND experiment and, on the theoretical side, the three neutrino model of Eq. (\ref{eq:3numodel}). ($\hat y_{23}\neq0$ gives a weaker improvement in the fit to KamLAND compared to $\hat y_{13}\neq0$ and, furthermore, might have additional implications.)

\subsection{KamLAND}
KamLAND is an experiment designed to be sensitive to $\Delta m^2_{21}$ and $\sin\theta_{12}$. The sources of the neutrinos are 53 nuclear reactors, with an average flux-weighted distance of 180 km to the detector. We study the sensitivity of KamLAND to the resonance effects by taking the following steps:
\begin{enumerate}
\item Calculate the expected number of unoscillated $e$-events using the fluxes of the nearest 21 reactors.
\item Calculate the energy dependent survival probability for each reactor for a set of values of $(m_\phi,\hat y_{13})$.
\item Calculate the expected number of oscillated $e$-events and obtaining the overall survival probability $P_{\rm calc}(E_i,m_\phi,\hat y_{13})$.
\item Perform a $\chi^2$ test,
\beq
\chi^2(m_\phi,\hat y_{13})=\sum_{i={\rm bins}}\left[\frac{P_{\rm calc}(E_i,m_\phi,\hat y_{13})-P_{\rm meas}(E_i)}{\sigma(E_i)}\right]^2,
\eeq
where $P_{\rm meas}(E_i)$ and $\sigma(E_i)$ are the experimental values provided by KamLAND \cite{KamLAND:2013rgu}. In our calculation, we assumed the unperturbed vacuum neutrino parameters to be $\theta_{12}=0.606\,,\theta_{23}=0.83\,, \theta_{13}=0.1\,, \Delta m^2_{21}=7.54\times 10^{-5}~\rm{eV}^2$ and $\Delta m^2_{32}=2.5\times 10^{-3}~\rm{eV}^2$. The standard matter propagation effects were taken into account by using the matter-modified mixing angle $\theta_{12 M}$ and mass splitting $\Delta m^2_{21 M}$ given in Eq. (3) and Eq. (4) of~\cite{KamLAND:2013rgu}.
\end{enumerate}
 
\begin{figure}[ht]
 \begin{center}
\includegraphics[width=0.5\textwidth]{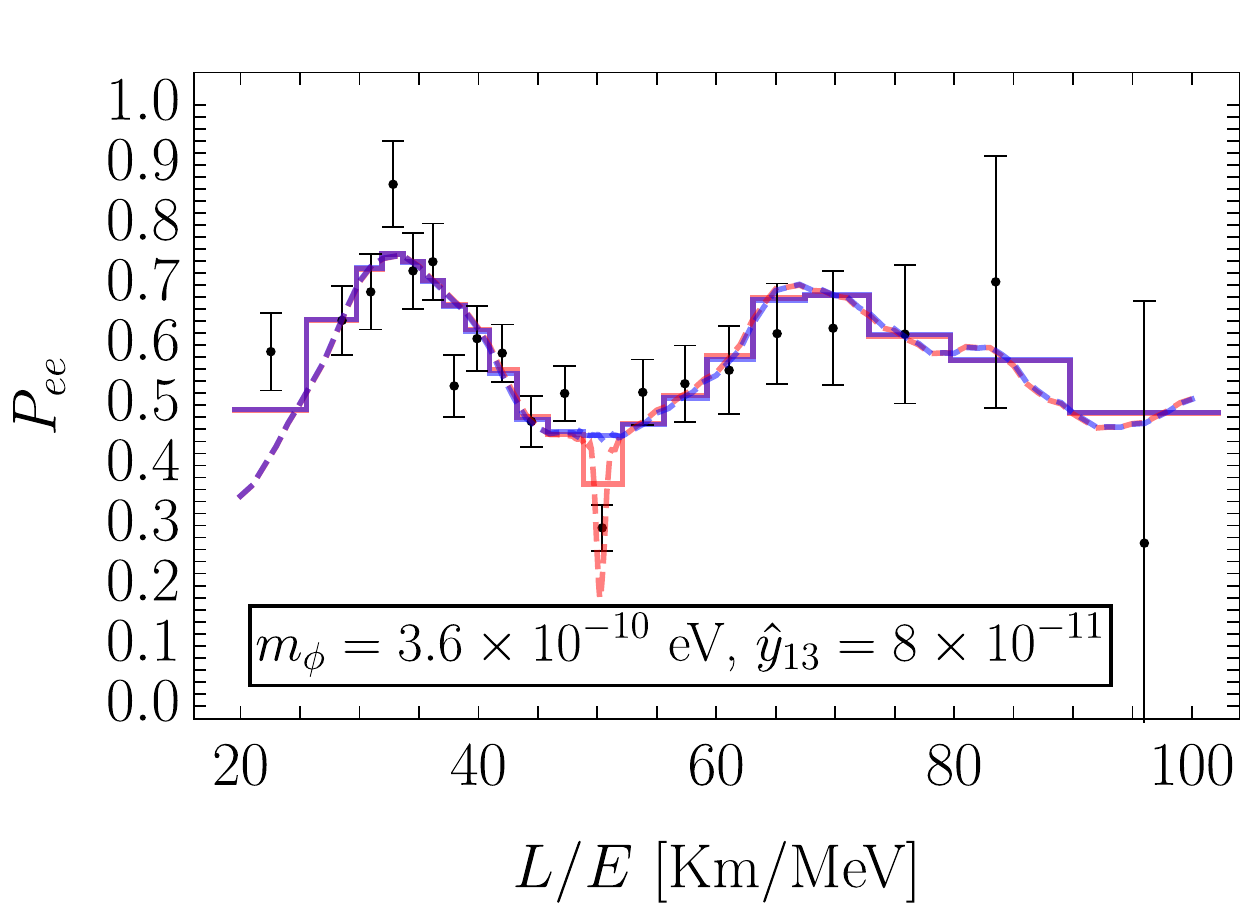}\hfill
\includegraphics[width=0.48\textwidth]{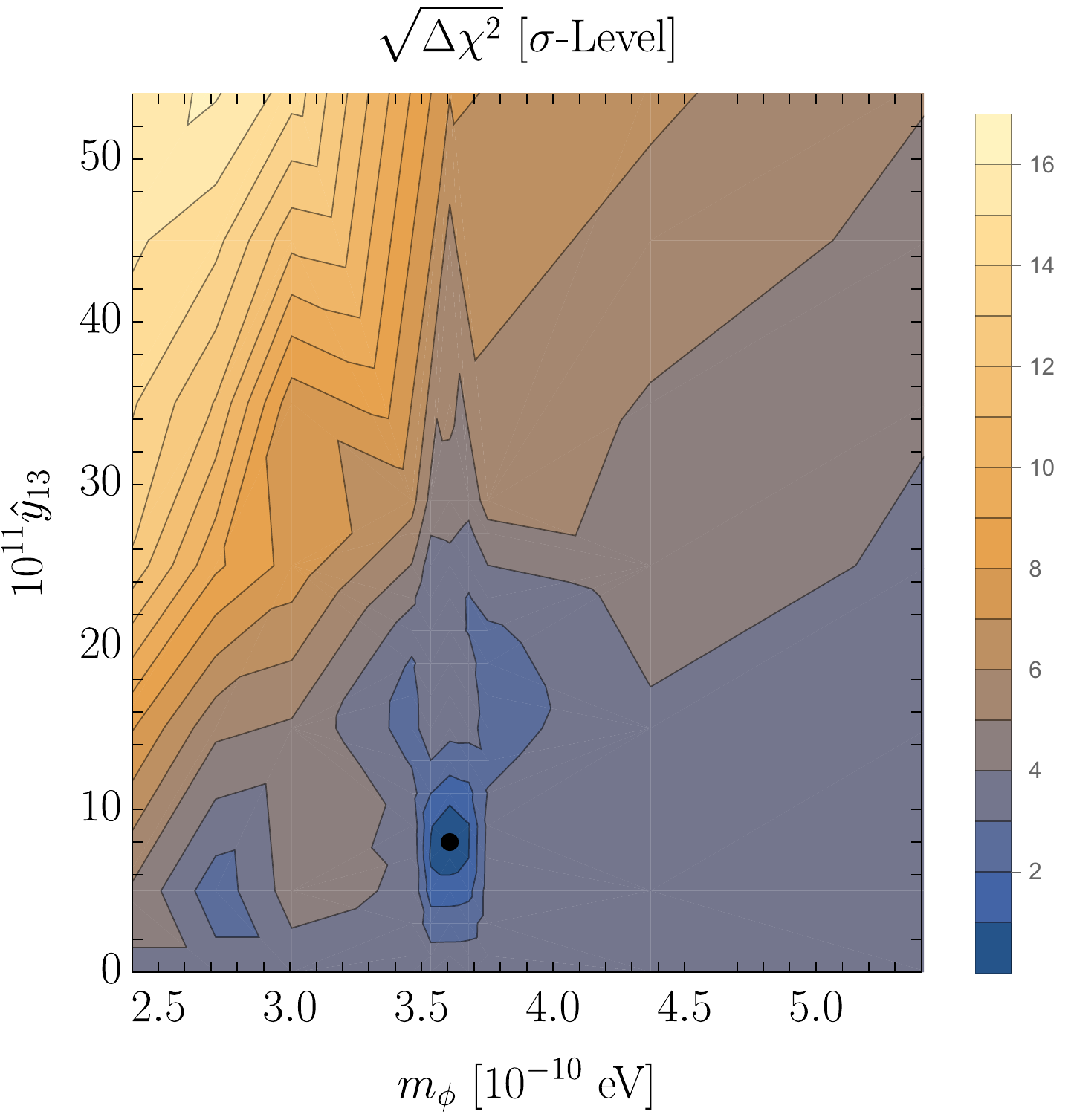}
  \caption{The KamLAND experiment: \textit{(Left)} $P_{ee}(L/E)$: The experimental results (black), the best fit of standard oscillations in matter (blue), and the best fit for a \ac{ULDM} field of $m_\phi=3.6\times 10^{-10}$ eV and $\hat y_{13}=8\times 10^{-11}$ (red). Shown are both the binned (solid) and the unbinned (dashed) probabilities. \textit{(Right)} $\sqrt{\Delta\chi^2}$ for a \ac{ULDM} field of mass $m_\phi$, and coupling $\hat y_{13}$\,, compared to the minimum, which corresponds to $m_\phi=3.6\times10^{-10}$ eV and $\hat y_{13}=8\times 10^{-11}$.
  }
  \label{fig:kamland_uldm}
 \end{center}
\end{figure}

Our results are presented in Fig.~\ref{fig:kamland_uldm}. 
The blue curve corresponds to standard neutrino oscillations in matter, produced using the fluxes and distances of the nearest 21 reactors (see~\cite{Losada:2021bxx}), and is consistent with KamLAND's result.
Including the ULDM, the best fit values of the parameters are
\beqa\label{eq:klbest}
m_\phi&=&3.6\times10^{-10}\ {\rm eV},\no\\
\hat y_{13}&=&8\times10^{-11}\ \frac{\sqrt{\Delta m^2_{31}}}{m_3+m_1}\sqrt{\frac{\rho_{\rm DM}}{\rho^\oplus_{\rm DM}}}\,,
\eeqa
where $\rho_{\rm DM}$ is the measured dark matter density from~\cite{Berlin:2016woy}. The most prominent effect is the suppression of $P_{ee}$ in the bin around $E_R\sim3.6\ {\rm MeV}$. This feature corresponds to the deviation near 50 km/MeV in Fig.~5 of the KamLAND paper \cite{KamLAND:2013rgu}.\footnote{In the journal version of Ref.~\cite{KamLAND:2013rgu}, the following statement is made concerning this bin: ``We have inspected the apparent deviation near 50 km/MeV for systematic effects, and find none; it is statistical in nature, and disappears with a different choice of binning". Indeed, the parametric resonance in the fit with the \ac{ULDM} field is narrow, $|\delta E|/E_R\sim0.02$, and the consequent deficit would not be apparent with a choice of larger bins.}
The fit using the above values for the parameters is approximately $3.5\sigma$ better than standard matter oscillations. The significance will, however, decrease when the look elsewhere effect \cite{Gross:2010qma} is taken into account.

Taking $m_1=0$ and $\rho^\oplus_{\rm DM}=\rho_{\rm DM}$, the best fit values correspond to
\beq
\epsilon_{31}=0.019,\ \ \ \epsilon_{31}m_\phi L/4=1.54.
\eeq
%

\subsection{JUNO}
\begin{figure}[ht]
 \begin{center}
\includegraphics[width=0.5\textwidth]{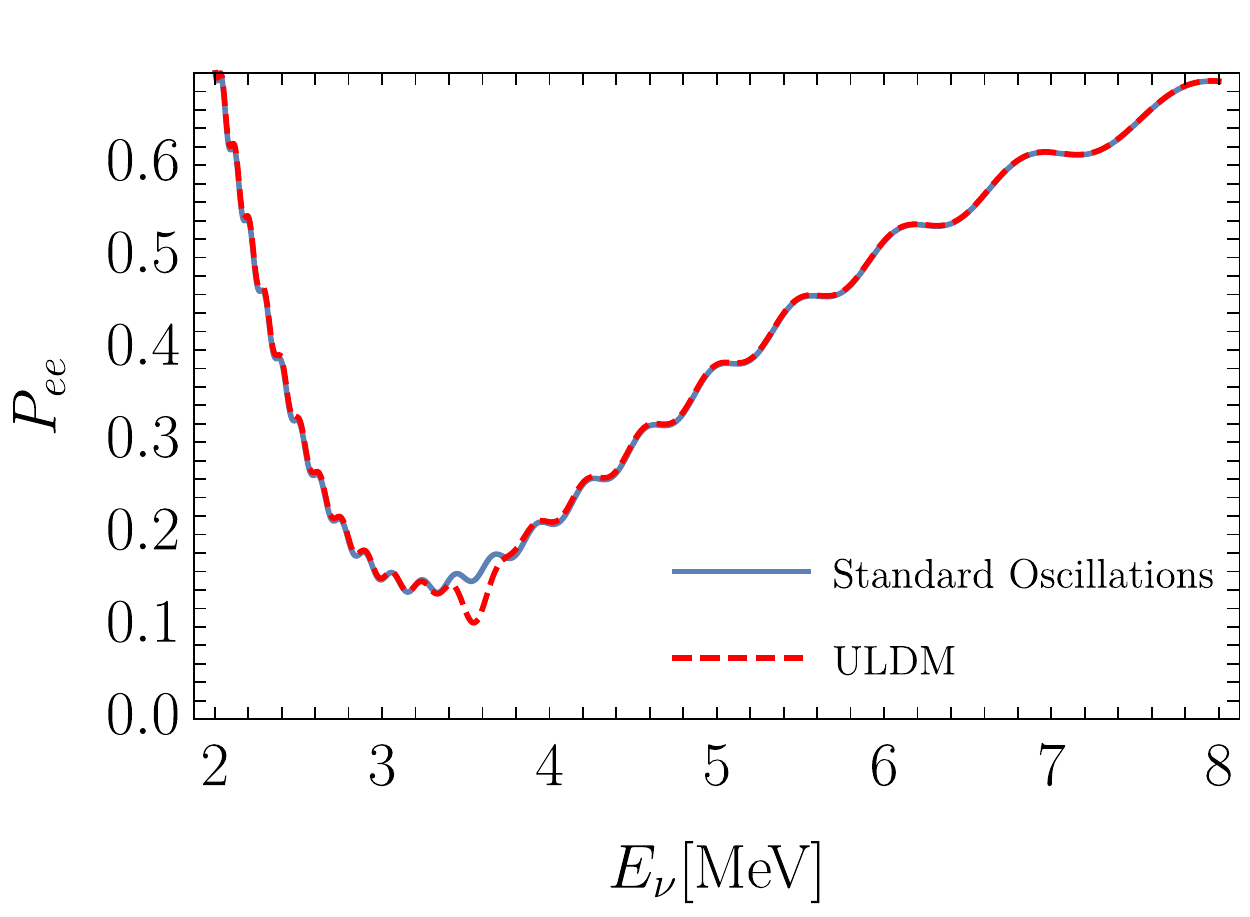}\hfill
\includegraphics[width=0.46\textwidth]{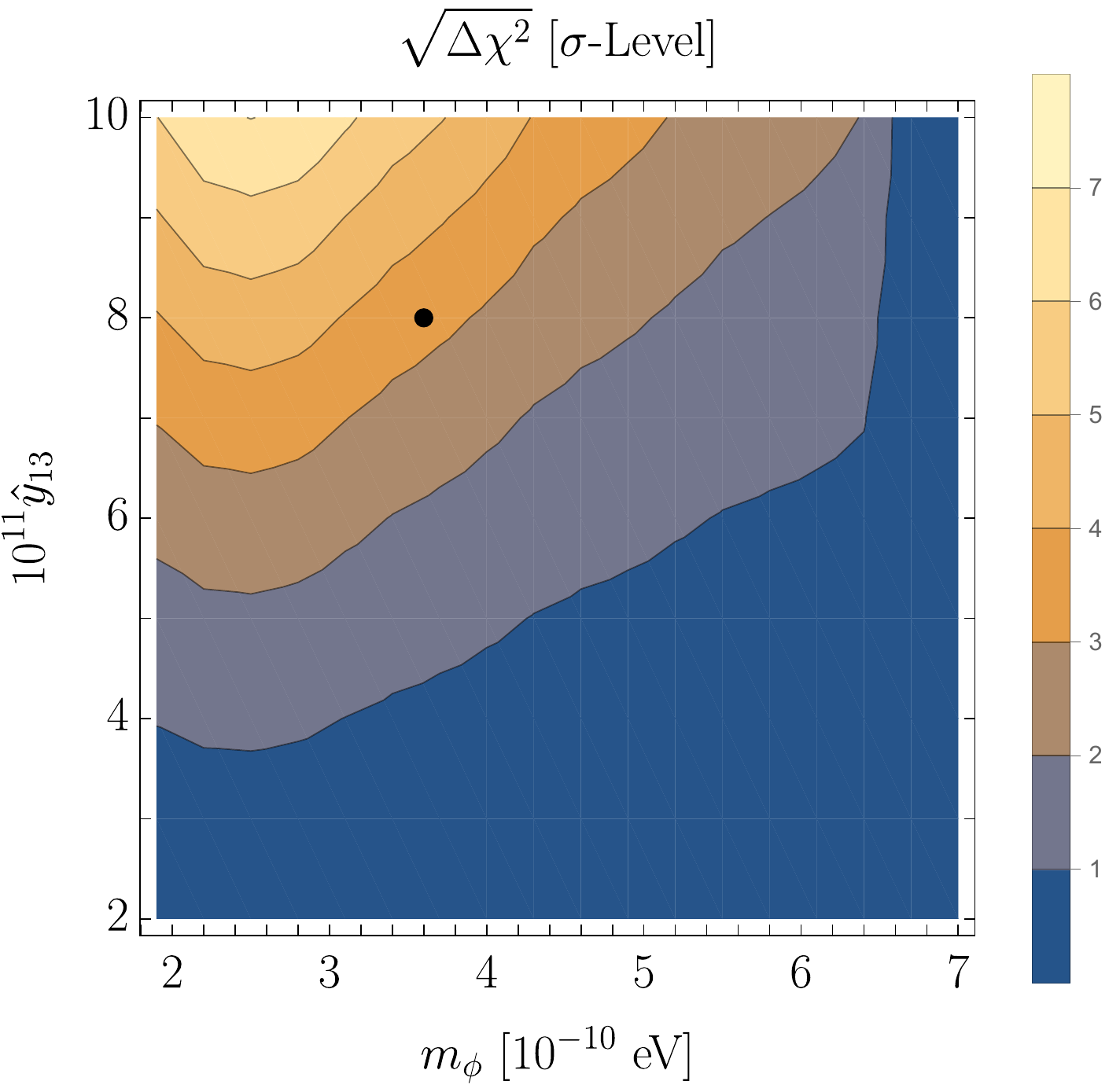}
  \caption{
  The JUNO experiment: \textit{(Left)} $P_{ee}(E)$: The expected result from standard oscillations (blue), and the expected result for a \ac{ULDM} field of $m_\phi=3.6\times 10^{-10}$\,eV and $\hat y_{13}=8\times 10^{-11}$ (red). \textit{(Right)}: Expected $\sqrt{\Delta\chi^2}$ between vacuum oscillations and oscillations in a \ac{ULDM} field, assuming 60 days of data taking. KamLAND best fit parameters are marked by a black dot.
  }
  \label{fig:juno_uldm}
 \end{center}
\end{figure}

JUNO is an experiment designed to be sensitive to both $\Delta m^2_{21}$ and $\Delta m^2_{31}$, in order to determine the neutrino mass hierarchy \cite{JUNO:2015zny}. The sources of the neutrinos are two reactors located at a similar distance of 53 km. We now study the effect that a \ac{ULDM} field with the parameters of Eq.~(\ref{eq:klbest}), corresponding to the best fit to the KamLAND measurements, would have on the survival probability measured by JUNO. The modification to the vacuum oscillation probability is shown in Fig.~\ref{fig:juno_uldm}. As expected, $P_{ee}$ is modified around $E_R\sim3.6\ {\rm MeV}$. 

The high statistics and good energy resolution of JUNO will enable the experiment to split the spectrum in the range of $2-8$ MeV into 200 energy bins. We performed a $\chi^2$ analysis to estimate the sensitivity of JUNO to the resonance effect, using
\beqa
\Delta \chi^2(m_\phi,\hat y_{13})&=&\sum_{i={\rm bins}}\left[\frac{P_{\rm uldm}(E_i,m_\phi,\hat y_{13})-P_{\rm vac}(E_i)}{\sigma(E_i)}\right]^2,\no\\
\sigma(E_i)&=&\sqrt{\frac{P(E_i)[1-P(E_i)]}{N(E_i)}},
\eeqa
where $N(E_i)$ is the expected number of unoscillated events in the $i$'th bin. 

Our results are presented in Fig.~\ref{fig:juno_uldm}. If the best fit values for KamLAND, Eq.~(\ref{eq:klbest}), are realized in nature, the deviation from the standard matter oscillation prediction will be signaled with a $3.5\sigma$ significance after $60$ days of collecting data. It is also important to note that JUNO could constrain other values of the UDLM parameters.

\subsection{Other experiments}
We studied whether a \ac{ULDM} field with the parameters of Eq.~(\ref{eq:klbest}) will modify the measurements of other neutrino experiments. Our conclusion, that currently only JUNO can test this scenario, is based on the following considerations:
\begin{itemize}
\item Solar neutrino experiments are unaffected because of our choice of $\hat y_{i2}=\hat y_{2i}=0$, meaning that the \ac{ULDM} field does not affect the propagation of $\nu_2$.
\item Atmospheric neutrino experiments are unaffected because $E^{\rm atm}\gsim 25E_R^{\rm kam}$ and $L^{\rm atm}=(0.03-250)L^{\rm kam}$. A resonant effect in both KamLAND and atmospheric neutrino experiments can only occur if the one at KamLAND arises from a high harmonic, $N\gsim25$. Eq. (\ref{eq:epsilonN}) implies then a highly suppressed $\epsilon_N$ at KamLAND, with baseline distance that is not large enough to generate a significant effect.
\item Accelerator neutrino experiments are unaffected because $E^{\rm acc}\gsim 10^3E_R^{\rm kam}$, and the same argument as for the atmospheric neutrinos holds.
\item The Daya Bay experiment, which has an energy spectrum similar to KamLAND, will not be sensitive to the $\phi$-field because of its much shorter baseline, $L^{\rm DB}\sim1\ {\rm km}$.
\end{itemize}

Of course, it could be that the deficit around $E\sim3.6\ {\rm MeV}$ in KamLAND would turn out to be a statistical fluctuation, and a \ac{ULDM} field exists with parameters different from  Eq.~(\ref{eq:klbest}). We leave a study of the sensitivity of neutrino experiments to other ranges of $\phi$-field parameters to future work.

\acresetall
\section{Conclusions}
\label{sec:con}
An \ac{ULDM} field $\phi$ with a mass $m_\phi$ greater than $\sim10^{-12}$ eV would oscillate with a period $\tau_\phi$ shorter than a few milliseconds. If it couples to neutrinos, this time scale is relevant to terrestrial neutrino propagation, which is observed over baselines $L$ from tens to thousands of km. We have shown that if the \ac{ULDM} oscillates many times during neutrino propagation {\it i.e.} $m_\phi L\gsim1$, a parametric resonance effect could be generated if a neutrino energy value of $E_R=\Delta m^2_{ij}/(2m_\phi)$ falls within the window of sensitivity for a given experiment.  We have further shown that the contribution of the \ac{ULDM} field  to the neutrino oscillation probability scales with $\sin^2\left(\epsilon_{ij} m_\phi L/4\right)$, where $\epsilon_{ij}$ 
denotes the ratio between the contribution of the \ac{ULDM} field to the neutrino mass to the vacuum value of the neutrino mass. This means that a resonant enhancement of the \ac{ULDM} effects would occur even for very small $\epsilon_{ij}$ values.

Given that $\Delta m^2_{31}\sim10^{-5}\ {\rm eV}^2$ and assuming $m_\phi\sim10^{-10}\ {\rm eV}$, the effect would be most relevant for $E\sim$ a few MeV. Indeed, we find that the long baseline reactor experiment KamLAND is sensitive to this type of effects. In fact, KamLAND has measured a significant deficit of events in one of its energy bins, around 3.6 MeV. Such a deficit, if not a statistical fluctuation, can be accounted for by a \ac{ULDM} field of mass $m_\phi\sim3.6\times10^{-10}\ {\rm eV}$, and with effective coupling to $\nu_1\nu_3$ of order $\hat y_{13}\sim10^{-10}$. This scenario improves the fit compared to standard matter oscillations with a local significance $\sim3.5\sigma$.

A future test of this scenario is expected from the medium baseline reactor experiment JUNO. A \ac{ULDM} field with parameters fitted to the KamLAND measurement is expected to give a $3.5\sigma$ deficit of events, compared to standard matter oscillations, in the energy bin around 3.6 MeV for 60 days of collected data.

We expect other neutrino experiments to be sensitive to other regions in the \ac{ULDM} parameter space ($m_\phi,\hat y_{ij}$). We leave a detailed study of these constraints to future work.     

\subsection*{Acknowledgements}
We thank M.C. Gonzalez Garcia for fruitful discussions regarding the KamLAND results. YN is the Amos de-Shalit chair of theoretical physics, and is supported by grants from the Israel Science Foundation (grant number 1124/20), the United States-Israel Binational Science Foundation (BSF), Jerusalem, Israel (grant number 2018257), by the Minerva Foundation (with funding from the Federal Ministry for Education and Research), and by the Yeda-Sela (YeS) Center for Basic Research. The work of GP is supported by grants from BSF-NSF, Friedrich Wilhelm Bessel research award, GIF, ISF, Minerva, SABRA - Yeda-Sela - WRC Program, the Estate of Emile Mimran, and The Maurice and Vivienne Wohl Endowment. IS~is supported by a fellowship from the Ariane de Rothschild Women Doctoral Program.

\appendix
\section{Resonance Equations of Motion for a Complex Coupling}\label{appendix:complex}
We consider the following Yukawa matrix in the neutrino mass basis in the two-generations picture:
\beq
\label{eq:2nuYMatrix-A}\hat y=y\begin{pmatrix} 0 & 1\cr 1 & 0 \end{pmatrix},
\eeq
where $y$ is complex. Omitting terms proportional to the unit matrix, which do not affect neutrino oscillations, we obtain
\beq
{\cal H}=2\omega_E[\sigma_z+\left(\epsilon^R_\phi\sigma_x+\epsilon^I_\phi\sigma_y\right)\sin(m_\phi t+\varphi)],
\eeq
where $\sigma$ are the Pauli matrices and
\begin{align}
\epsilon^R_\phi\equiv \frac{2\phi_0\text{Re}\left(y\right)}{m_1-m_2} \qquad & \,, \qquad \epsilon^I_\phi\equiv \frac{2\phi_0\text{Im}\left(y\right)}{m_1+m_2}\,.
\end{align}
Following the same prescription from the main text, Eqs.~\eqref{eq:nu0sol}-\eqref{eq:coom}, we find the following equations for $c_{1,2}$,
\beq\label{eq:coomcomplex}
i\partial_t c_{1,2}=\frac{\omega_E}{2i}\left(\epsilon^R_\phi\pm i\epsilon^I_\phi\right)\left[e^{i(m_\phi\pm2\omega_E)t+i\varphi}-e^{-i(m_\phi\mp2\omega_E)t-i\varphi}\right]c_{2,1}\,,
\eeq
which in the rotating wave approximation, keeping only the slowly oscillating term, yield
\beq
i\partial_t c_{1,2}=\pm i\frac{\omega_E}{2} r_\epsilon e^{\mp i\left(m_\phi\delta_E t +\varphi-\varphi_\epsilon\right)} c_{2,1}\,,
\eeq
where $r_\epsilon \equiv\sqrt{{\epsilon^R_\phi}^2+{\epsilon^I_\phi}^2}$ and  $\varphi_\epsilon\equiv \arctan\left(\epsilon^I_\phi/\epsilon^R_\phi\right)$. Therefore, in the proximity of the resonance, the phase $\varphi_\epsilon$ can be absorbed into the \ac{ULDM} phase $\varphi$, and will thus have no observable effect in the time-averaged case.



\end{document}